\documentclass[preprint,12pt]{elsarticle}

\usepackage{amssymb}

\usepackage[hyphens]{url}
\usepackage{hyperref}
\usepackage{color,soul}
\sethlcolor{white}

\journal{Journal of Information Security and Applications}

\begin{document}

\begin{frontmatter}



\title{Denial of Wallet - Defining a Looming Threat to Serverless Computing}


\author{Daniel Kelly\corref{cor1}}
\ead{d.kelly69@nuigalway.ie}
\author{Frank G. Glavin}
\ead{frank.glavin@nuigalway.ie}
\author{Enda Barrett}
\ead{enda.barrett@nuigalway.ie}
\cortext[cor1]{Corresponding author}
\address{School of Computer Science,\\ National University of Ireland, Galway (NUIG),\\ Galway, Ireland}

\begin{abstract}
Serverless computing is the latest paradigm in cloud computing, offering a framework for the development of event driven, pay-as-you-go functions in a highly scalable environment. While these traits offer a powerful new development paradigm, they have also given rise to a new form of cyber-attack known as Denial of Wallet (forced financial exhaustion). In this work, we define and identify the threat of Denial of Wallet and its potential attack patterns. Also, we demonstrate how this new form of attack can potentially circumvent existing mitigation systems developed for a similar style of attack, Denial of Service. Our goal is twofold. Firstly, we will provide a concise and informative overview of this emerging attack paradigm. Secondly, we propose this paper as a starting point to enable researchers and service providers to create effective mitigation strategies. We include some simulated experiments to highlight the potential financial damage that such attacks can cause and the creation of an isolated test bed for continued safe research on these attacks. 
\end{abstract}




\end{frontmatter}



\section{Introduction}
\label{introduction}
Serverless computing is an application deployment architecture that aims to provide pay-as-you go event driven functionality. Applications are developed as per each desired process and the event that invokes it. Serverless function platforms provide the infrastructure to deploy code for execution across their cloud and define the event processing logic that prompts the functions to run using the model: \emph{event, trigger, and action}. Serverless computing abstracts back-end management from users, allowing only minimal access to some basic parameters such as function memory allocation and function run time timeout. Functions execute on the platform's traditional Infrastructure as a Service (IaaS) virtual machine (VM) offerings, however the provision of such VMs is managed by the platform in response to function invocation and not by the developer. Upon these execution VMs, a function container is spun up where the serverless function will execute. A container will be be made for each function invocation and containers may be reused for repeat invocations. Unlike IaaS, you do not pay for the uptime and resources consumed by the ``execution'' VM or function container but rather for the run time of each function, hence the name Function as a Service (FaaS). This system allows for serverless applications to scale massively as the application is hosted on the serverless platform's cloud and as such has access to its resources. However, this has led to a threat unique to serverless computing, a new form of cyber-attack called Denial of Wallet (DoW). It may operate in a similar fashion to Denial of Service (DoS) attacks, where a bad actor initiates a barrage of requests to a service. In DoS, this would use up available resources rendering the service unusable. Although, as serverless applications are incredibly scalable, more and more functions will be invoked and quickly the application owner will incur a large bill exhausting their finances, hence the term Denial of Wallet attack.

We perform an evaluation of the damage that could be incurred by a DoW attack via cost analysis of the four major commercial serverless platforms; AWS Lambda\footnote{\url{https://aws.amazon.com/lambda/}}, Google Cloud Functions\footnote{\url{https://cloud.google.com/functions}}, Microsoft Azure Function\footnote{\url{https://azure.microsoft.com/en-us/services/functions/}} and IBM Cloud Functions \footnote{\url{https://cloud.ibm.com/functions/}}, in various scenarios. We also devise a means of testing DoW attacks in a safe manner that would not result in real financial exhaustion by creating a serverless platform emulation.

\section{Related Work}
\label{related}
To date, there has been no academic analysis of DoW attacks. A number of blogs have drawn attention to the issue \cite{Dooley2019, Sachenko2020, Rahic2018} although they only go as far as to describe what DoW is. Pursec \cite{Pursec2018} released a list of ten security risks that face serverless computing with DoW featuring on that list. They do not go into detail on DoW specifically, instead focusing on DoS attacks and suggesting mitigation techniques such as:
\begin{enumerate}
\item Writing efficient serverless functions, which perform discrete targeted tasks.
\item Setting appropriate timeout limits for serverless function execution.
\item Setting appropriate disk usage limits for serverless functions.
\item Applying request throttling on API calls. 
\item Enforcing proper access controls to serverless functions. 
\item Using APIs, modules and libraries which are not vulnerable to application layer DoS attacks such as ReDoS and Billion-Laughs-Attack
\end{enumerate}

While these points serve as a starting point for protection against DoW attacks, they are not a solution. Writing efficient serverless functions will of course minimise functions with excessive run times. However, this does not mitigate against potential attacks spanning long timescales and high volumes of attacking nodes, that will inevitably drive the function run time up. Setting limits such as timeouts, disk usage and API throttling, shifts the attack from DoW (exhausting finances), to DoS (exhausting resources). Access control is only as effective as its difficulty in acquiring credentials. While administrator specific functions may be safe, API endpoints accessible to account holders of an application are susceptible to exploitation by fake users i.e. bots or ``Sybils'' (discussed later in Section \ref{fakes}).

\hl{Since no DoW attacks have been publicly recorded, at the time of writing, we are in a rare position of being ahead of the attackers. We can preemptively develop mitigation strategies rather than reacting to an attack when it occurs. As such, we must first theorise the mechanisms that malicious entities may use. We can achieve this be drawing inspiration from other well studied attacks.}
Application layer DoS attacks are the strongest contender for the starting point of effective DoW attacks. As mentioned in point 6 above, such attacks could be re-purposed for DoW use, potentially leaving APIs, modules and libraries vulnerable to such modified attacks.

Further related work that is valuable to the analysis of DoW attacks are on serverless security and DoS mitigation.\hl{ Since we are ahead of any known large scale attacks, it is important to utilise all knowledge from related fields}. Datta \emph{et al.} \cite{Datta2020} proposes a system for mitigating indirect function flow by monitoring the activity of every function API trigger. This research was directed at preventing information theft. However, analysis of all exposed API Gateways is necessary for DoW mitigation as any unprotected function triggers will serve as a point of attack. Inspiration for DoW mitigation may be taken from research carried out by DoS mitigation. Barna \emph{et al.} \cite{Barna2012} proposed a system for DoS mitigation that dynamically changes the rules on a firewall to minimise false positive detections and then pass suspicious activity to a CAPTCHA challenge. The application layer was the area of the web application being monitored. This makes this mitigation system relevant to DoW research, as DoW attacks are conducted on the application layer via API triggers. Their experimental setup of a mock application that they then attacked with real DoS tools served as inspiration for the experimental setup devised in this paper (see Section \ref{testbed}). As DoW attacks only aim to exhaust finances, slow rate attacks may be preferable as they are more difficult to detect as observed by Mukhopadhyay \emph{et al.} \cite{Mukhopadhyay2010}.

\section{Defining Denial of Wallet Attacks}
\label{definition}
We define DoW attacks as the intentional mass, and continual, invocation of serverless functions, resulting in financial exhaustion of the victim in the form of inflated usage bills. Execution costs are increased via excessive function run time, invocation count and additional resource consumption. This is achieved by exploiting the massive capability for scaling, as the platform will handle the increased load but will incur costs in doing so. 

A DoW attack is a form of attack unique to serverless computing. Given the scaling capabilities of serverless platforms, as a result of being hosted on a vast cloud of virtual machines, the effects of a regular DoS attack can be dealt with by massively scaling to handle the volume of requests. However, this scaling comes at a price. Serverless, or Function-as-a-Service, bills the application owner per run time of each function invocation. The resulting volume of function invocations following a DoS attack would therefore incur financial exhaustion of the application owner. \hl{As such, the term DoW came to describe this attack since unlike DoS, where the service is targeted and disrupted, the ``wallets'' (that is the finances) are the target. The result is an application owner who must now pay for function executions that are not in fact \textit{legitimate} usage.} 

One of the earliest uses of the phrase in a publication is in the Open Web Application Security Project's (OWASP) report on the security risks of serverless computing \cite{OWASP}. It is mentioned as a risk to consider along with traditional DoS attacks. Prior to this, DoW was often referred to as a ``financial exhaustion attack'', ``denial of capital'' or simply ``serverless denial of service''. However, we believe that a ``serverless DoS'' attack and a DoW attack can be seen as unique threats, with the former focusing on serverless specific resource exhaustion and the latter being as we have described.

\subsection{Preemptive Examination of a Threat}
\label{scope}
\hl{
To date, there is little to no targeted research on DoW. We believe this is due to an absence of publicly known occurrences of attacks. This poses a unique opportunity to be ahead of attackers by mitigating against attack methods that are early in development. The vulnerability exists in serverless technology and it would be na{\"i}ve to believe it will not be exploited. In Section {\ref{motivation}} we discuss the potential motivations to conduct such an attack and potential scenarios they may be executed in.

Research on DoW will encompass strategically approaching the issue from the perspective of a bad actor and is an example of preemptive cybersecurity investigation. This domain shares many key attributes with bot detection, DoS mitigation and cloud security, meaning it also inherits their pitfalls and difficulties.

This research aims to bring a dormant threat in the conscience of the community and potentially address it before it may even begin and develop effective mitigation strategies before it can become a prominent threat.
}

\section{Mechanisms of a Potential Attack and  Mitigation Strategies}
\label{attack}

Since there have been no reported major instances of DoW, it is necessary to theorise attack strategies that may be developed. We take precedent from the techniques employed by DoS attacks and analyse their potential for modification as a DoW attack. Much of the strategies used in DoS attacks are viable in a DoW setting. However, we propose a number of unique cases that invariably distinguish DoW as its own potential threat.

\subsection{Traditional Attack Methods}
As serverless computing operates on the ``No Ops'' principle (that is the developer is not involved with the setup or maintenance of the application's back-end), popular traditional DoS attack patterns such as ICMP Flood, SYN Flood and UDP Flood\cite{Douligeris2004} may not necessarily work as functions themselves and do not possess the infrastructure such attacks target. If the goal is to target serverless functions specifically rather than taking down an application as a whole, the most appropriate attack method would be HTTP flooding given that the majority of serverless functions work from API triggers.
Two potential DoS attacks that could be re-purposed for DoW are:
\\
\\
\textbf{ReDoS - }Regular expression Denial of Service (ReDoS) \cite{OWASPa} is an algorithmic complexity attack that produces a DoS by providing a regular expression (regex) that takes a very long time to evaluate. The attack exploits the fact that most regular expression implementations have exponential time worst case complexity,  so for larger input strings (the ‘evil regex’), the time taken by a regex engine to find a match increases exponentially. The aim of the attacker is to provide such regex(s) so that it takes an indefinite amount of computation time which in turn will either slow down the application or completely bring it down.
\\In a DoW context, this would drive the run time of the function as high as possible. Serverless functions have a timeout parameter that is an upper limit most commonly of five minutes. However, with functions being billed per 100 ms, this time limit is capable of inflicting notable financial damage.
\\
\\
\textbf{Billion-Laughs-Attack - }This is a DoS attack that targets parsers of XML documents \cite{Morgan2014}. The attack is also known as an XML bomb or Exponential Entity Expansion  Attack (XEE Attack). An example attack consists of defining 10 entities, each defined as consisting of 10 of the previous entity, with the document consisting of a single instance of the largest entity, which expands to one billion copies of the first entity.
\\Again, this is another potential method in which the run time of a function can be driven up to cause a DoW.
\\
\\
These traditional attack methods serve as inspiration for one aspect of potential DoW attacks, that is, driving up the function run time. In order to maximise the effectiveness of such methods, they should be used in conjunction with some mechanism of increasing the number of function invocations. 

\subsection{Short Time Span Attack}
Short time span attacks are any form of ``flood'' attack. As many requests as possible are sent to the server as quickly as possible. This is the basis of DoS.
A Web Application Firewall (WAF) is the recommended mitigation approach for large spike HTTP flooding attacks. An effective strategy against HTTP flooding DoS attacks is rate limiting requests from a single IP address via the WAF.
Commercial platforms offer WAF services that implement safety against OWASP's top 10 security risks \cite{OWASP} and provide rate limiting rules such as AWS\footnote{\url{https://aws.amazon.com/waf/}}, Google Cloud\footnote{\url{https://cloud.google.com/armor}}, and Microsoft Azure\footnote{\url{https://azure.microsoft.com/en-us/services/web-application-firewall/}}. These rules are difficult to set. A hard limit will protect against short time span (large spike in requests) HTTP flooding but at the risk of producing false positives in the event of actual real traffic spikes. Adaptive rules could be deployed but would become a target for long term false normative baseline inducing attacks (Section \ref{fnb}).

Rate limiting also exists within the API Gateway services themselves. These are generally limits that show off the service's ability to handle large traffic (AWS API Gateway can handle 10000 requests per second\cite{AWS}). Unlike rate limiting rules on the firewall, if the API rate limit is reached, this could incur a DoS result as resources are taken up. Efforts to stop the effect of DoW by limiting the massive scaling factor of serverless applications could instead cause DoS as limits are reached.

\subsection{Long Time Span Attack}
\label{long}
There are a number of ``slow'' DoS attacks, as described by Zargar \emph{et al.}\cite{Zargar2013}, such as Slowloris and R-U-Dead-Yet (RUDY). These attacks, through various means, aim to send high workload requests that lock up connection resources i.e. sockets. Analysis on such attacks has proved they are a threat \cite{Tripathi2018}.

A new strategy may be employed for long term DoW attacks that differs from the previous interpretation of ``slow'' attacks. We propose the idea of ``leeching'' as a potential form of DoW. A leech being a malicious program that continually triggers API endpoints that invoke functions. These leeches operate indefinitely and at a rate that would not be detected by reasonable WAF rate limiting rules. The key difference between DoW and DoS is that DoW does not need to execute all at once to consume all available resources to cause damage. As such, a leech DoW attack could be executed over a long time span and appear as legitimate traffic. We demonstrate the potential damage a leech DoW attack could create in Section \ref{costanalysis}

\subsection{Distributed Denial of Wallet}
A long time span attack would be the most simple method of bypassing WAF rules and implementing a DoW attack. However, a single attacker or ``leech'' may not inflict financial damage on a scale that would affect larger organisations.
Multiple leeches could be deployed onto a botnet either willingly, such as through IRC chat as seen in Operation Payback \cite{Sauter2013}, or via malware that makes unknowing victims hosts for the leeches.
A distributed approach to DoW would ultimately prove difficult to detect given how closely it would resemble real traffic using a long time span attack protocol.

\subsection{Adaptive Mitigation Deception}
\label{fnb}
In DoS mitigation systems that use an adaptive rule set i.e. can change the firewall rules base on fluctuation in ``deemed to be real'' traffic \cite{Barna2012}, a DoW attack can again utilise the fact that it can be executed over a long time span. Unlike a rigid firewall rule where it will be forever limited, if the attacker gently increases the intensity in such a way that the mitigation system adapts to it, it will then establish new false normative baselines.

\subsection{Serverless Exploitation}
\label{exploit}
As well as API triggers, serverless functions can also be invoked through uploading to a storage service (such as S3 buckets for AWS Lambda). This introduces the concept of serverless specific attack patterns such as function input parameter exploitation, by flooding a storage service with images of varying size depending on the limits imposed by the function.

To fully realise the potential harm a DoW attack may be capable of, it should not be simply seen as a slow burning DoS attack. Instead it should fully exploit any oversights the developer may have made. If there is no limit imposed upon functions that process files, the run time of a function will increase. We show this effect in Section \ref{exploittest}.

Similarly on Microsoft Azure Functions, the developer does not configure the memory allocation of a function unlike its competitors. This could similarly be abused by forcing a function to scale to a higher memory allocation to cope with larger inputs. 

\subsection{Fake Users}
\label{fakes}
A novel approach that could be taken towards performing a DoW attack would be the mass generation of fake users on a user subscription web application. Varol \emph{et al.}\cite{Varol2017} found that between 9\% and 15\% of Twitter accounts are bot accounts. That equates to roughly 49.5 million accounts based off the current monthly active user statistics \cite{SalmanAslam2020}. Using a tool such as Selenium\footnote{\url{https://www.selenium.dev/}}, which automates browser activity, a mass of clients could perform a site walkthrough creating many fake users. Not only would this trigger many functions involved with profile creation but also, as discussed in Section \ref{exploit}, uploading the largest images possible (as profile pictures) can be used to increase run time. A further result of the mass generation of fake users would be access to other API endpoints such as page querying or other site functionality further falsely invoking more functions causing more damage.

In recent studies, there have been three main approaches to fake user detection: graph-based, crowd sourcing and machine learning.  

Graph-based detection is the use of graphs constructed from the social network of users to understand the network information and the relationships between edges or links across accounts to detect bot activity. There are many approaches to graph-based bot detection, a popular one being the use of ``random walks'', for example SybilRank \cite{Cao2012}, Criminal  account  Inference  Algorithm  (CIA) \cite{Yang2012} and SybilWalk \cite{Jia2017}. SybilWalk claims a more robust classification with 99\% of the top 80,000 nodes ranked in order of likeliness to be a Sybil, being correctly classified Sybils. However, in the ranking lists produced by SybilRank and CIA, only 0.3\% and 30\% are Sybils, respectively. Accounts are ranked by performing a random walk based movement on an undirected social graph. The idea of the random walk method is to label human users with benignness scores and Sybil users with badness scores. The score is then used in ranking the users.

Crowd sourcing is the use of humans to label whether a given user is a Sybil or a human. They identify, evaluate, and determine behaviors that would point towards an account being a bot. Alarifi \emph{et al.} \cite{Alarifi2016} recruited 10 volunteers deemed to have expert knowledge of Twitter (all BSc Computer Science graduates and active Twitter users) to rate and label 2000 random accounts captured via a Tweet streaming API. Using this, a ground truth data set was established for the further training and comparison of various machine learning algorithms such as C4.5 and Random Forest.

The machine learning method involves developing algorithms and statistical methods that can develop an understanding of the revealing features or behavior of social network accounts in order to distinguish between human- and computer-led activity. Ersahin \emph{et al.} \cite{Ersahin2017} used a Na\"ive Bayes based classification algorithm on a data set that had undergone preprocessing via Entropy Minimization Discretization (EMD). The results were 86.1\% correct classification before EMD and 90.9\% after.

\section{Denial of Wallet Scenarios}
\subsection{Attack Motivation}
\label{motivation}
\hl{Motivations for attacks such as DoS are largely the same; some interested party wishes to cause a disruption to a victim's service as a form of revenge, ransom, threat, retaliation or simply a desire to cause chaos. DoW is different in this regard. however, due to our proposed ``slow-burn'' and deceptive attack methods, we believe that the motivations to cause such damage are slightly more nuanced. We akin the motivation to perform DoW on a victim to that of espionage or sabotage. One party wishes to covertly effect another in a negative way thus giving the attacking party the upper hand when a critical matter arises. 

One example we propose that best exemplifies this is that of a party wishing to stifle the growth of a startup company. A startup company offers some service powered by serverless computing. This service has potential to pull clientele from the attacking party. The attacking party may then initiate a DoW ``leech'' that will gradually ramp up fictitious usage of the startup's service. This will drive up the startup's serverless usage bill for no tangible return, unbeknownst to them. The startup will continue to pay for their serverless platform and may also invest in other areas given their new found usage. The attacker then ceases the DoW reducing the startup's usage to a much lower level that does not justify the capital invested. The startup finds themselves in a position where they have spent a large amount of their budget on the serverless platform's bills and may have taken loans backed by their previous fake success that they cannot pay. This will cause the stifling or even shut down of the startup.

This example can be applied to any situation where a smaller entity is vulnerable to their serverless powered service being targeted for DoW. Given that the attack's most powerful trait is their ability to go unnoticed, it could be used as a tool for larger politically interested entities to silence or weaken any opposition. Fpr example, finances spent on inflated serverless bills is money not spent on organising demonstrations etc.}

\subsection{Attack Execution}
For DoW to occur, a malicious entity must gain access to function endpoints. This can be broadly achieved via knowledge that certain services use serverless functions e.g. Seattle times image loading\footnote{\url{https://aws.amazon.com/solutions/case-studies/the-seattle-times/}}. More targeted approaches could be taken to extract specific endpoints of functions via web scrapers looking for direct API links that follow the typical naming convention e.g.

AWS - https://xxxxx.execute-api.xx-xxxx-x.amazonaws.com/xxxxx/xxxxx

Azure - https://xxxxx.azurewebsites.net/api/xxxxx 

GCF - https://xx-xxxxx-spherical-plane-xxxxx.cloudfunctions.net/xxxxx\\ 

It is good practice amongst industry professionals not to leave API URLs visible like this, however, such a vulnerability could still exist through human error.\\ 

The core mechanism of a DoW attack will be to spam function triggers. We consider two likely scenarios for this to occur.\\

\textbf{Scenario 1: }Function API endpoint URLs have been scraped or leaked. These URLs are loaded into spamming software that continuously hits these endpoints with HTTP requests. The rate of attack can vary depending on the security of the application. In the presence of no DDoS security (rate limiting etc.), fast paced attacks can take place but will raise suspicion when an application owner sees a sudden spike. Slow rate attacks, ``Leeches'', are effective for DoW due to their difficulty of detection. They are akin to ``Bad Bot Traffic'', fictitious traffic generated by bot nets intended to give false analytics. As such, slow rate DoW has a double impact of driving up operational costs with function invocations and spoofing application owners with falsified traffic (potentially causing further financial damage when used to justify increased spending).\\

\textbf{Scenario 2: }Developers have successfully hidden or obscured API URL endpoints or the endpoint is a different trigger such as upload to a storage system. Attacks will need to be performed as a fake application user. Web application testing software such as Selenium\footnote{\url{https://www.selenium.dev/}}  can automate interactions with a web application. Applying a similar tactic to a bot net will produce fictitious traffic as with the URL example. The downside to this is that it requires greater computing resources and knowledge of which applications use serverless functions and how they are triggered.  

\section{Isolation Zone for Denial of Wallet Attack Testing}
\label{testbed}
\begin{figure}[!t]
\centering
\includegraphics[width=\linewidth]{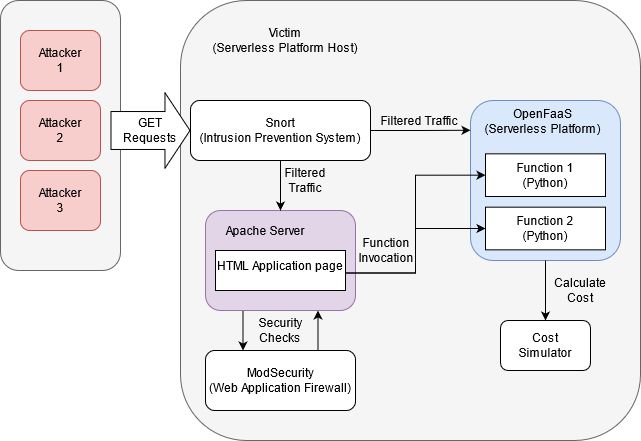}
\caption{Test bed of an isolated serverless platform for attack simulation.}
\label{testbedfig1}
\end{figure}

\hl{
As DoW is a financial attack, it is impractical to test such on real commercial platforms. Given that there are no known historical examples of DoW, we must have a means of generating data for training of mitigation systems and then validating such systems. To facilitate this, an isolated serverless platform was devised that will emulate the cost damage of DoW. This consisted of four desktop PCs running in the same network. One machine was designated the ``victim''. It would play host to the serverless platform and cost emulator being targeted by our attacks. The other three machines were designated as ``attackers''. It is from these machines that we will launch a variety of attack strategies to determine the best candidacy for DoW. The topology of our test bed is illustrated in Figure {\ref{testbedfig1}}.}

\subsection{Configuration}
\hl{
The ``victim'' machine is a Linux desktop running the Ubuntu 18.04 operating system. The serverless platform used to execute functions is OpenFaaS\footnote{\url{https://www.openfaas.com/}}. OpenFaaS is a framework for building serverless functions with Docker and Kubernetes which has first class support for metrics. Any process can be packaged as a function enabling you to consume a range of web events without repetitive boiler-plate coding. The mock application that will trigger the functions on OpenFaaS runs on Apache Server 2\footnote{\url{https://httpd.apache.org/}}. The Apache HTTP Server Project is an effort to develop and maintain an open-source HTTP server for modern operating systems including UNIX and Windows. In order to perform more meaningful experiments on DoW, our environment has a twofold security system. Where commercial platforms offer one Web Application Firewall (WAF) that can run on all its products e.g. AWS WAF\footnote{\url{https://aws.amazon.com/waf/}}, finding a single catch-all solution to secure both our application server and serverless platform proved difficult. Our solution was to secure the Apache server with the ModSecurity v2\footnote{\url{https://modsecurity.org/}} WAF and use the Intrusion Prevention System (IPS) Snort v2\footnote{\url{https://www.snort.org/}} to perform WAF actions on the OpenFaaS deployment.}

\subsection{Serverless Deployment}
\label{openfaas}
\hl{
OpenFaaS was chosen as the serverless platform for executing function code. Some of its highlights include;}
\begin{itemize}
\item{\hl{Ease of use through UI portal and one-click install}}
\item{\hl{Write functions in any language for Linux or Windows and package in Docker/OCI image format}}
\item{\hl{Portable - runs on existing hardware or public/private cloud with Kubernetes or containerd}}
\item{\hl{CLI available with YAML format for templating and defining functions}}
\item{\hl{Auto-scales as demand increases}}
\end{itemize}
\hl{OpenFaaS runs on the ``PLONK'' stack. PLONK is a Cloud Native stack for building applications which stands for:}
\begin{itemize}
\item{\hl{Prometheus - metrics and time-series}}
\item{\hl{Linux - OS or service mesh}}
\item{\hl{OpenFaaS - management and auto-scaling of compute - PaaS/FaaS, a developer-friendly abstraction on top of Kubernetes. Each function or microservice is built as an immutable Docker container or OCI-format image.}}
\item{\hl{NATS - asynchronous message bus / queue}}
\item{\hl{Kubernetes - declarative, extensible, scale-out, self-healing clustering}}
\end{itemize}
\hl{The function gateway can be accessed through its REST API, via the CLI or through the UI. All services or functions get a default route exposed, but custom domains can also be used for each endpoint. Prometheus collects metrics which are available via the function gateway's API and which are used for auto-scaling.
Functions are invoked from the application front end via an API gateway. This also exposes the functions to direct invocation (as is normal in FaaS). The front end and OpenFaaS communicate via HTTP requests and JSON.}
\subsection{Web Application Server Deployment}
\label{apache}
\hl{The Apache server hosts the application front end. It is a powerful, flexible, HTTP/1.1 compliant web server that
implements the latest protocols, including HTTP/1.1 (RFC2616). It is highly configurable and extensible with third-party modules, which we make use of when implementing our security solution. Apache runs on Windows 2000, Netware 5.x and above, OS/2, and most versions of Unix, as well as several other operating systems, meaning we could easily pair it with our OpenFaaS deployment on Ubuntu 18.04.
This server will be used to host mock applications for future experiments.}
\subsection{Security}
\label{security}
\hl{
As previously mentioned, where commercial platforms offer a single WAF that protects all their products, this proved unrealistic for our environment. We split the security measures to 1. protect the application server and 2. protect the serverless platform.

The application server is protected by ModSecurity. It is a toolkit for real-time web application monitoring, logging, and access control. Its capabilities include:}
\begin{itemize}
    \item {\hl{Real-time application security monitoring and access control}}
    \item {\hl{Full HTTP traffic logging}}
    \item {\hl{Continuous passive security assessment}}
    \item {\hl{Web application hardening}}
\end{itemize}
\hl{
We use this a the primary means of securing the application from attacks such as XML or Shell injection etc.

A highly recommended means of mitigating HTTP Flood attacks (which is the core of worst case scenario DoW attacks) is rate limiting based on IP address. For this, all inbound traffic gets filtered through Snort. Snort is an open source network intrusion detection system (NIDS). Snort is a packet sniffer that monitors network traffic in real time, scrutinizing each packet closely to detect a dangerous payload or suspicious anomalies.
Snort is based on libpcap (for library packet capture), a tool that is widely used in TCP/IP traffic sniffers and analyzers. Through protocol analysis and content searching and matching, Snort detects attack methods, including DoS, buffer overflow, CGI attacks, stealth port scans, and SMB probes. Running in ``inline'' mode, Snort becomes an intrusion prevention system. Inline mode is where Snort sits between two bridged network interfaces, analysing network packets as they come in. When suspicious behavior is detected, it will drop any packets that trigger certain rules. Otherwise it will pass them to the Apache server or OpenFaaS back end (depending on the function invocation method).}

\subsection{Serverless Platform Pricing Emulator}
\label{emulator}
\hl{
We developed the Serverless Platform Pricing Emulator in order to compare the effects of DoW across multiple commercial platforms. It is a program that sits on top of an OpenFaaS serverless deployment that allows you to emulate how much its function executions would cost on AWS Lambda, Google Cloud Functions, IBM Cloud Functions and Microsoft Azure Functions. It is written in Python and converts usage metrics gathered by Prometheus on the OpenFaaS platform to a comparable cost had those functions been invoked on the four largest commercial serverless platforms. The cost calculator queries Prometheus for the total number of successful function invocations and the cumulative execution time. As all functions are configured to 128MB, delineation between function executions is not required and a total cost can be calculated as per each platform's pricing guidelines. The source code\footnote{\url{https://github.com/psykodan/openfaas-commercial-platform-emulator}} is available for use by the community who wish to perform similar experiments on serverless computing financial effects.}

\subsection{Attacking Nodes}
\label{attackers}
\hl{
The attack machines execute a python script that sends multiple GET requests to the functions and change the IP address of the machine (simulating any number of attack nodes required). 
For DoW, slow rate attacks are the primary concern, however, DoS style of attacks may also be executed. 

Attacker may also be configured with the Kali Linux operating system\footnote{\url{https://www.kali.org/}} for use of existing attack software. In Section {\ref{demonstrationdow}}, we do so and utilise the DoS tool Low Orbit Ion Cannon.}
\section{Demonstrating Denial of Wallet}
\label{demonstrationdow}
\begin{figure}[!t]
\centering
\includegraphics[width=\linewidth]{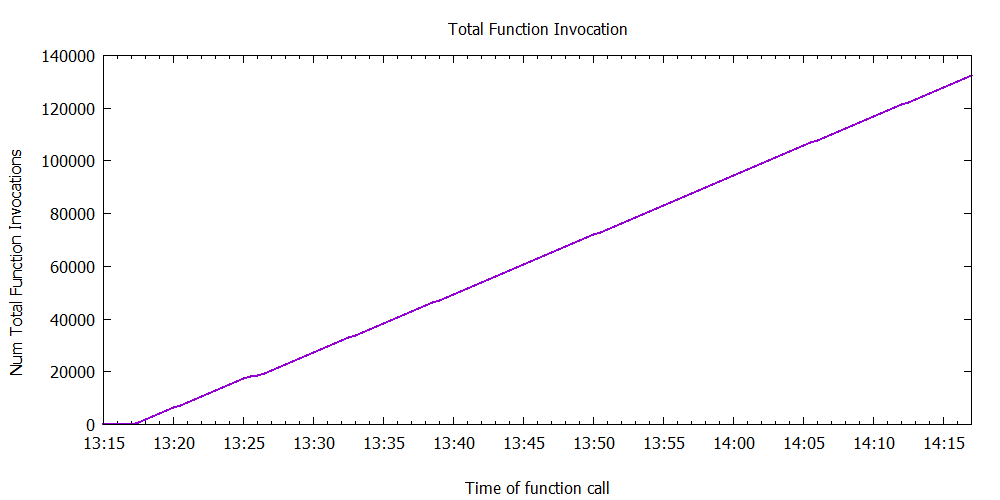}
\caption{Total function invocations gathered from HTTP flood attack with no protection. }
\label{1hrTotal}
\end{figure}

\begin{figure}[!t]
\centering
\includegraphics[width=\linewidth]{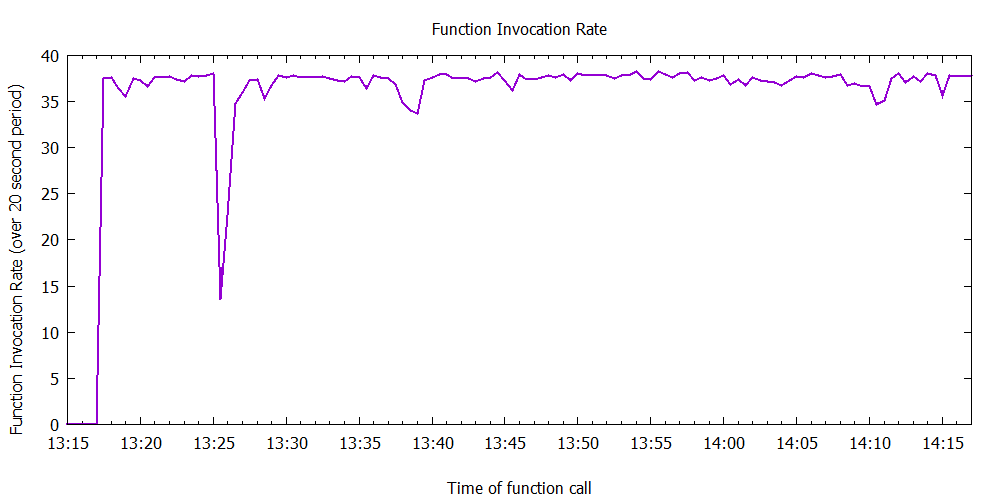}
\caption{Function invocation rate per 20 second interval gathered from HTTP flood attack with no protection. }
\label{1hrRate}
\end{figure}

\begin{figure}[!t]
\centering
\includegraphics[width=\linewidth]{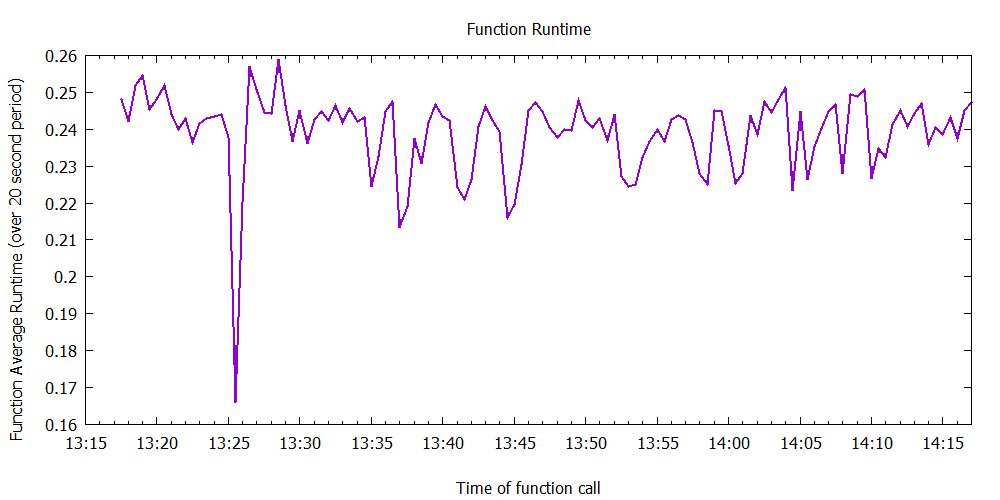}
\caption{Average function run time per 20 second interval gathered from HTTP flood attack with no protection. }
\label{1hrRuntime}
\end{figure}

\begin{figure}[!t]
\centering
\includegraphics[width=\linewidth]{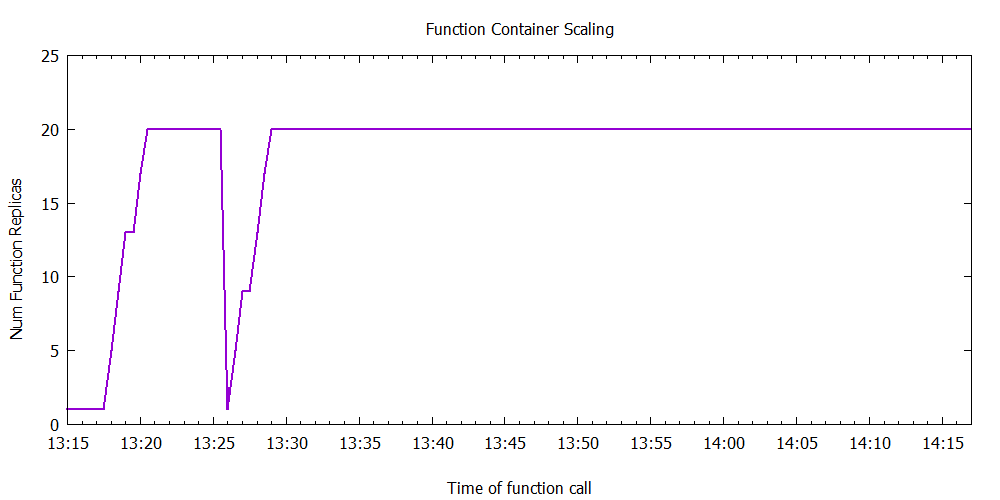}
\caption{Metrics gathered from HTTP flood attack with no protection. Function replica generation}
\label{1hrReplica}
\end{figure}

Using our test bed, we can demonstrate the mechanism of DoW attacks. Firstly, we utilise an existing DoS tool, Low Orbit Ion Cannon (LOIC), a DoS tool made famous by the ``Operation Payback'' attacks in defence of Wikileaks in 2010, that can perform HTTP flooding attacks {\cite{Roman2011}}.
We configured a single node with Kali Linux and ran LOIC with a single serverless function on our OpenFaas platform as the target.\hl{ We allowed LOIC to execute for one hour, continuously spamming direct function invocations to our OpenFaaS deployment.  Our platform gathers metrics such as function invocation rate, container scaling, total invocations and execution time using Prometheus and Grafana. We observed that over 130,000 function invocations could be triggered in this time span (Figure {\ref{1hrTotal}}). Average invocation rate and run time per 20 second interval is also recorded(Figures {\ref{1hrRate}} and {\ref{1hrRuntime}}). This run demonstrated how a serverless platform scales to accommodate large volumes of requests by creating additional containers for function execution (Figure {\ref{1hrReplica}})}
However, this example is a worst case scenario where a single attacker is allowed to continuously spam requests. A Web Application Firewall (WAF) can mitigate simple GET request flooding using rate limiting. Using LOIC again and activating our intrusion prevention system (IPS) that acts as a WAF to our OpenFaaS deployment, we can limit the damage any one node can do with an HTTP Flooding attack by rate limiting requests from a single source. This would suggest that short time span, single node attacks are not suitable for DoW.

From this demonstration, we can infer that an unprotected serverless function is susceptible to DoW from a single node. Implementing rate limiting protects against the issue. However, as previously discussed in Section \ref{long}, DoW can be executed over a much longer time span than a DoS attack, thus rate limiting will be ineffective on such cases.  

\section{Theoretical Damage Analysis}
\subsection{Serverless Function Input Parameter Exploitation}
\label{exploittest}
A common use case of serverless functions is in image re-processing, such as creating a thumbnail from a user uploaded image \cite{Akiwatkar}. We use AWS Lambda as an example to demonstrate how one could exploit input parameters to a serverless function by uploading large images that must be resized to a thumbnail size. For this, an AWS Lambda function was created that resizes images uploaded to an AWS S3 bucket to a $128 \times 128$ px thumbnail \cite{Kumar2019}. The thumbnail image is then stored in its own separate bucket. Images of increasing size were uploaded and we observe the effect on function run time. The function memory allocation was kept constant at 512MB as we found lower memory allocations could not handle the largest images. Results are shown in Table \ref{table1}. 

As expected, there is an increase in run time with an increase of image size. In this scenario, it would fall to the developer to impose a hard limit on image size. However, as smartphones are producing higher resolution images with each new flagship device, for example, the baseline of an acceptable maximum image size gets pushed higher. This makes even well thought out functions susceptible to input exploitation.

\begin{table}[!t]
\renewcommand{\arraystretch}{1.0}
\caption{Effect on run time of increasing image size}
\label{table1}
\centering
\begin{tabular}{|c|c|c|c|c|c|c|}
\hline
Image size & Test 1 & Test 2& Test 3& Test 4& Test 5& Average\\
$l \times w$ (px) & (ms)&(ms)&(ms)&(ms)&(ms)&(ms)\\
\hline
540	& 248.27 & 230.51 & 254.77&	239.32&	236.22&	241.818\\
\hline
1080&371.18&368.43&361.15&385.22&409.24&379.044\\
\hline
2160&1045.59&1051.62&1083.57&1046.05&1059.11&1057.188\\
\hline
4320&2945.17&2958.46&2954.22&2963.07&2933.42&2950.868\\
\hline
8640&9901.85&9843.55&9924.92&9877.68&9881.59&9885.918\\
\hline
\end{tabular}
\end{table}

\subsection{Cost Analysis of Excessive Function Invocation}
\label{costanalysis}
\begin{figure}[!t]
\centering
\includegraphics[width=\linewidth]{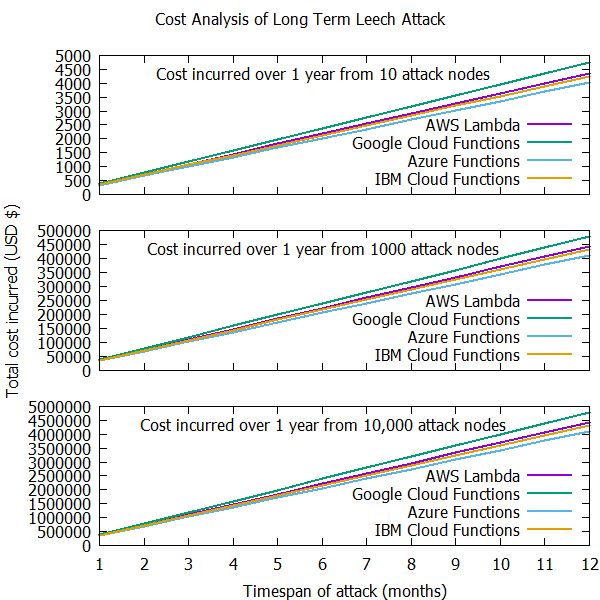}
\caption{Cost incurred when an increasing number of nodes send 2000 requests every hour}
\label{attack}
\end{figure}
As a reference, an Apple iPhone X takes photos producing an image of $4032 \times 3024$ px \cite{Lim2018}. Cross-referencing our prior experiments on input exploitation (Section \ref{exploittest}), we can take a run time of 2950ms of a memory allocation of 512MB as a benchmark for a legitimate function that a malicious user would consider targeting for a Denial-of-Wallet attack. 

A cost analysis of this function on the four largest commercial serverless platforms; AWS Lambda, Google Cloud Functions, Microsoft Azure Functions and IBM Cloud Functions, was performed for increasing number of function invocations using the Serverless function cost calculator we devised as the mock application on our isolated test bed (Section \ref{testbed}).

This calculator takes the number of function requests, the run time of each of those functions and the memory allocation of the function as inputs and returns the total cost of executing those functions inclusive of all free allowances granted by each platform and additional costs such as API call costs. The official cost guides for each platform were used; AWS Lambda\footnote{\url{https://aws.amazon.com/lambda/pricing/}}, Google Cloud Functions\footnote{\url{https://cloud.google.com/functions/pricing/}}, Microsoft Azure Function\footnote{\url{https://azure.microsoft.com/en-us/pricing/details/functions/}} and IBM Cloud Functions \footnote{\url{https://cloud.ibm.com/functions/learn/pricing}}.

The three graphs show a linear increase of cost incurred with Google Cloud functions resulting in the greatest charges followed by AWS Lambda, IBM Cloud Functions and lastly Azure Functions. With a modest bot-net of 1000 nodes, a slow rate attack of 2000 requests per hour will cost an application owner roughly \$40,000 after one month and between \$400,000 and \$500,000 if left unchecked for a year. A bot-net of 10,000 nodes will do the same damage in one month that 1000 nodes would do in a year.

\section{Discussion}
\label{discussion}
\hl{
DoW is a strong potential threat, as shown by our preliminary experiments on the subject. It's a rare opportunity to be ahead of a threat, as such it is necessary to momentarily become malicious and devise how these attacks may be conducted; ``If you know the enemy and know yourself, you need not fear the result of a hundred battles''.
}
Using previously designed attacks and our own research on the potential damage of DoW we can theorise the following would lead to an effective attack:
\begin{itemize}
  \item Use Sybils to interact with a serverless application in order to trigger its functions.
  \item Do not aim to hit an application with DoS levels of requests i.e. HTTP Flooding.
  \item Use slow rate/long term attacks to naturally raise the invocation count of functions.
  \item Make traffic look as natural as possible, the aim is to fool an application owner into believing they have many users. Not only does this cause damage via DoW over serverless but the additional damage caused by falsely believing in a larger user base resulting in additional spending.
  \item Super-long term attacks yield high damage but are susceptible to discovery. Aim between one and six months for an attack. Discovery is more likely to occur when finances are being analysed.
\end{itemize}
\hl{Mitigation against such attacks will prove difficult. We believe approaching it from the same mindset as bot detection rather than DoS will be appropriate. Some potential avenues for consideration are:}
\begin{itemize}
  \item \hl{Constructing common usage graphs to be used in traffic analysis for flagging suspicious users}
    \item \hl{Strategic honey-pots that only malicious web-scrapers would find}
      \item \hl{Applying reinforcement learning to graphs generated by user interaction to determine how suspicious they are}
\end{itemize}

\hl{The value of the damage inflicted need not be astronomical to be effective. As mentioned in the motivations for DoW attacks (Section {\ref{motivation}}) the most likely targets are smaller entities. The results in Figure {\ref{attack}} highlight that a modest bot-net of 1000 nodes could inflict up to \$50,000 of costs in a month. This would cause significant finance issues to a small startup company.}
\section{Conclusions and Future Work}
\label{conclusions}
From this initial investigation on the emerging threat of DoW attacks, we have concluded that there is potential for significant financial damage should a malicious entity wish to exploit unique selling points of serverless computing. 
Future work on the mitigation of such attacks will need to tackle the following concerns:
\begin{enumerate}
    \item Further research on fictitious request classification.
    \item Securing serverless applications from our proposed threat of ``leeches''.
    \item Further investigations into unique weaknesses of serverless computing for unique attacks.
    \item The detection of fake users and mitigation of their actions.
\end{enumerate}
Meaningful data must be gathered in the absence of historical attacks, rather than wait for large scale attacks to occur. The barrier to obtaining such data is the creation of realistic test rigs that can be put under various attack simulations without incurring real world financial damage. Such a barrier is what makes research of this topic currently more difficult to test than traditional cybersecurity research topics, where a test bed can be created in isolation with relative ease. Our implementation of an isolated serverless platform to emulate the billing practices of the four major commercial platform is a substantial step towards furthering research on this topic.

We conclude that effective DoW attacks will certainly be of a distributed nature. A single source is not capable of delivering severe damage in realistic scenarios where proper security procedures have been followed. It is then in conjunction with existing and future Distributed Denial of Service research that Distributed Denial of Wallet will be investigated.

In this paper, we propose that DoW attacks are one of the latest threats in cybersecurity. With the ever increasing adoption of serverless computing, it is a threat that must be considered. Our work has provided a baseline for further research to be carried out in the field. The simulation experiments we performed highlight the financial damage that could occur as a result of DoW. Our simple attacks incurred theoretical damage of over one million dollars. While these numbers seem outlandish, it is an important baseline for worst case scenarios that can be built upon to prevent such attacks becoming a reality.


\appendix

\bibliographystyle{elsarticle-num} 
\bibliography{dow.bib}






\end{document}